\documentclass[twocolumn]{revtex4} 
\usepackage{amsmath, amsthm, amscd, amssymb}
\usepackage{bm}
\usepackage{bbm}
\usepackage{graphicx}  
\usepackage{dcolumn}   
\usepackage{bm}        
\usepackage{amssymb}   
\usepackage{subfigure}

\let\vec\boldvec%

\begin{document}

\title{Stability of Chiral States, Role of Intermolecular Interactions and \\Molecular Parity Violation}

\author{Mohammad Bahrami}
\email{mbahrami@mehr.sharif.edu} \affiliation{Department of Chemistry, Sharif University of Technology, P.O.Box 11365-9516, Tehran, Iran }

\author{Afshin Shafiee}
\email{shafiee@sharif.edu} \affiliation{Department of Chemistry, Sharif University of Technology, P.O.Box 11365-9516, Tehran, Iran }

\begin{abstract}
 We study the problem of stability of chiral states, also known as problem of chirality, within the framework of a two-dimensional approximation of a symmetric double-well potential. We show how the symmetry breaking of the potential due to the molecular parity violation can stop the tunneling in a coherent way, accordingly stabilize the chiral states. Then, we use the quantum Brownian motion within a linear Lindblad-type equation to model how the intermolecular interactions make the tunneling incoherent, thus inducing a racemization by dephasing. Finally, we investigate the normal physical conditions where the dephasing racemization does not suppress the effects of the molecular parity violation, accordingly the molecular parity violation may be observed experimentally.
\end{abstract}
\maketitle

\section{Introduction}
Chirality is a fundamental concept in different branches of science.
The synthesis of chiral molecules in enantiopure forms, in particular for pharmaceutical uses, is a great challenge in the synthetic and structural chemistry~\cite{chi1,chi2,chi3}. In biology, perhaps the most outstanding unsolved problem is the origin of bimolecular homochirality \cite{biochi1,biochi3}. Chirality is also connected to the molecular parity violation where the weak interactions result in tiny energy differences between two chiral enantiomers~\cite{B,Dar,Q1,Let,Q2}.
The chiral behavior of a molecule can be described by a particle of mass $M$ moving in a double-well potential $V(q)$ with two minima at $q=\pm q_{0}/2$, separated by a barrier $V_{0}$, as shown in Fig.1; here $q$ is a generalized inversion coordinate \cite{Town,Her,LW}. In the absence of weak interactions, $V(q)$ is symmetric and so the corresponding eigenstates of Hamiltonian have definite parity and are delocalized in space. However, the chiral states are parity asymmetric and thus  cannot be stationary states. The existence of delocalized eigenstates may contradict the usual chemical view that considers the molecules as objects with a well-defined spatial structure. There are also strong evidences suggesting the existence of stable chiral states, e.g., many chiral molecules show stable optical activity which means the chiral states are stable.  Accordingly, although the delocalized states of some molecules clearly confirm some experimental results (e.g., inversion frequency of ammonia~\cite{Town}), for some other molecules there should be a mechanism to stabilize the chiral states. This leads to the problem of the stability of optical activity, known also as Hund's paradox or the problem of chirality \cite{hund,Mer}. Indeed, in the literature two different versions of chirality problem have been formulated. One is the original concern first expressed by Hund himself \cite{hund,Mer}: why {\it some} chiral molecules show a stable optical activity, associated to a well-defined chiral state, in spite of the fact that these states are not eigenstates of the Hamiltonian, in particular none of them corresponds to the ground state. In recent years, this initial concern seems to have been transformed into a stronger one~\cite{Cin,HS,SH,Sim,Pf,JZ,Jona,Ver,Ama,Wig,TH,Ber,AK}: why chiral molecules have a definite chirality. The new version of the question has been extended to all non-planar molecules, chiral or non-chiral, since their potential energy surface has at least two minima~\cite{Town, Her}. This has led to a very lively debate.

For an isolated molecule such as ammonia, the hydrogens beat between two minima of $V(q)$ with the frequency $\omega _{x}$. Using standard WKB methods, $\omega_{x}$ can be calculated as $\omega_{x} = A \: \omega_{0} (q_{0}/\sqrt{\hbar/M\omega_{0}}) \exp\left[-B \: V_{0}/\omega_{0}\right]$,
where $A$ and $B$ are constants, of order unity, which depend on the precise mathematical form of $V\left(q\right)$ \cite{LW}. Obviously, the slower the beating (i.e., tunneling through the barrier), the longer the life time of an isolated localized molecule, \textit{if} the molecule was initially prepared in a handed form. Accordingly, \textit{once produced}, the chiral states of molecules with high inversion barrier are stable for a remarkably long time, as explained by Hund \cite{hund,Mer}. However, upon a quantitative test, Hund's explanation seems rather insufficient for the molecules with low inversion barrier. Nor can Hund's answer explain why the habitual states of a molecule are the left-handed or right-handed isomers and not its parity eigenstates. After Hund, several attempts have been made to explain the problem by either taking into the account a small parity-violating term in molecular Hamiltonian \cite{Q1,Q2,HS,SH}, or by adding non-linear and stochastic terms to the dynamics because of interactions with a surrounding environment \cite{HS,SH,Sim,Pf,JZ,Jona,Ver,Ama,Wig,TH,AK,Ber}.

The molecular parity-violating energies (although not yet observed experimentally) may have possibly important consequences for the question of bio-homochirality \cite{Q1,Q2}. They can also play a promising role in the understanding and testing of the fundamental symmetries of physics \cite{Q1,Let,Q2,HR,BQ}. The molecular parity-violating energies has been also considered as a possible way to resolve the problem of optical activity \cite{Q2,HS,SH}. Here we focus on the connection of parity-violation and the problem of optical activity. We investigate this issue both for an isolated molecule and also for the one under the influence of environmental interactions.

\section{Quantum description of chiral molecules}.
For all non-planar molecules (chiral or achiral), the potential energy surface has at least two minima~\cite{Town,Her} in which the corresponding nuclei configuration states (i.e., chiral states) are localized. Thus, the chiral molecule, in its simplest form, can be described by a one-dimensional Hamiltonian with a double-well potential $V\left(q\right)$. In the absence of weak interactions, $V\left(q\right)$ is symmetric, and the energy eigenstates are also parity eigenstates:
\begin{equation}
\hat{H} \: \psi_{n}\left(q\right)=E_{n} \: \psi_{n}\left(q\right),
\quad
\psi_{n}\left(q\right)=\left(-1\right)^{n+1}\psi_{n}\left(-q\right).
\end{equation}
with $n=1,2,3,...$.
\begin{figure}
\begin{center}
{\includegraphics[scale=.7]{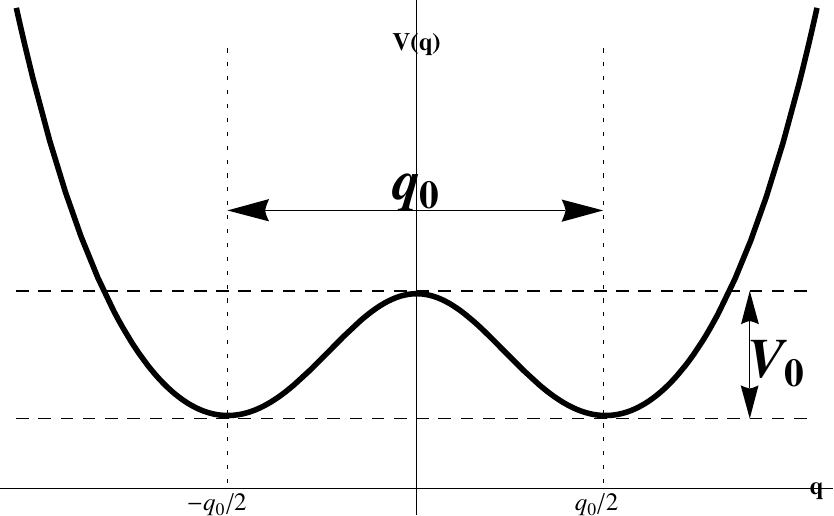}}
\caption{
A double- well potential provides a primitive but very constructive and convenient description of chiral molecules. The characteristics of the double-well potential, in particular the height of barrier ($V_0$) and the separation of two minima ($q_0$), can be determined from the spectroscopic data. The chiral states are localized at each minima.
}
\end{center}
\end{figure}
Taking into account the typical energy spectrum of $\hat{H}$, at the ordinary temperatures and at thermodynamic equilibrium, the fraction of molecules in the two lowest states, $\psi_{1,2}(q)$, are very close to unity (see Fig.2). However, $\psi_{1,2}(q)$ are delocalized between the two minima of $V\left(q\right)$, thus they cannot represent the chiral states of the molecule. The chiral states, which are localized each at one minimum, are then coherent superposition of $\psi_{1,2}\left(q\right)$:
\begin{equation}
\psi_{L}=\frac{1}{\sqrt{2}}\left[\psi_{1}+\psi_{2}\right],
\qquad
\psi_{R}=\frac{1}{\sqrt{2}}\left[\psi_{1}-\psi_{2}\right].
\end{equation}
The set $\left\{\psi_{L},\psi_{R},\psi_{n\geq3}\right\}$ provides the physically natural basis for the Hilbert space of our problem. Using this basis, $\hat{H}$ takes a block-diagonal form since its matrix elements $H_{ij}=\left\langle i\right| \hat{H} \left|j\right\rangle$ are:
\begin{equation}
H_{Ri} = H_{iR} = H_{Li}= H_{iL} = 0,
\quad
H_{ij} = E_{i} \; \delta_{i,j}
\label{eq:eq3}
\end{equation}
where $i,j \geq 3$. We denote the $2\times2$ square matrix of $\hat{H}$ in the subspace of $\left\{\psi_{L},\psi_{R}\right\}$ by $\hat{H}_{0}$. The matrix $\hat{H}_{0}$ is ``self-contained'', in the sense that the time evolution of $\psi_{L}, \psi_{R}$ depends only on $\hat{H}_{0}$. In other words, because of Eq.~\eqref{eq:eq3}, $\hat{H}_{0}$ is sufficient for describing the behavior of the chiral molecule, when only the first two eigenstates are thermodynamically relevant (i.e., when $V_0\gg k_BT$ with $k_B$ the Boltzmann constant and $T$ the temperature). The matrix elements of $\hat{H}_{0}$ can be written as
\begin{equation}
\hat{H}_{0}=\omega _{0}\:\hat{I}+\vec{\omega}.\vec{\sigma}
\end{equation}
with $\vec{\omega}=\left(\omega_{x},0,\omega_{z}\right)$, $\vec{\sigma}=\left(\hat{\sigma}_{x},\hat{\sigma}_{y},\hat{\sigma}_{z}\right)$ the Pauli matrices, and:
\begin{eqnarray}
\omega_{0}&=&\frac{1}{2}\left(E_1 + E_2\right),
\quad
\omega_{x}=\frac{1}{2}\left(E_{1}-E_{2}\right),
\\
\omega_{z}&=&\frac{1}{2}\left(H_{LL}-H_{RR}\right).
\end{eqnarray}
where $\omega_{z}$ is the measure of molecular parity violations.
The largest numerically calculated value of $\omega_z$ is of the order of $3\times10^{-2}$Hz (e.g., for S$_2$Cl$_2$)~\cite{Q2,BQ}. Although parity-violation effects are very small (compare it with $V_0$ which is in the range $10^{10}-10^{15}$Hz \cite{Town,Her}) nevertheless when $\omega_z\gg\omega_x$ they dramatically change the eigenstates of $\hat{H}_0$, from non-localized states $\psi_{1,2}$ to localized states $\psi_{L,R}$, which can be considered as a partial solution of the problem of optical activity.

However, a chiral molecule is not an isolated system and it interacts constantly with its environment, e.g., through the collision with the background gas. Meanwhile, one of the main challenges in  observing the molecular parity violation is how to get rid of intermolecular collisions~\cite{Dar}. Since the parity violating terms are so small (at best, $\omega_z$ is of the order of few Hertz~\cite{Dar}) the question may arise that in what experimental conditions (pressure and temperature) the molecular parity violation still determine the dynamics of chiral molecule and so may be observed experimentally. Our approach is different from the usual view manifesting effects of intermolecular collisions in frequency shift and broadening of the molecular lines. In the following, we discuss our position in detail.

\begin{figure}
\begin{center}
{\includegraphics[scale=.9]{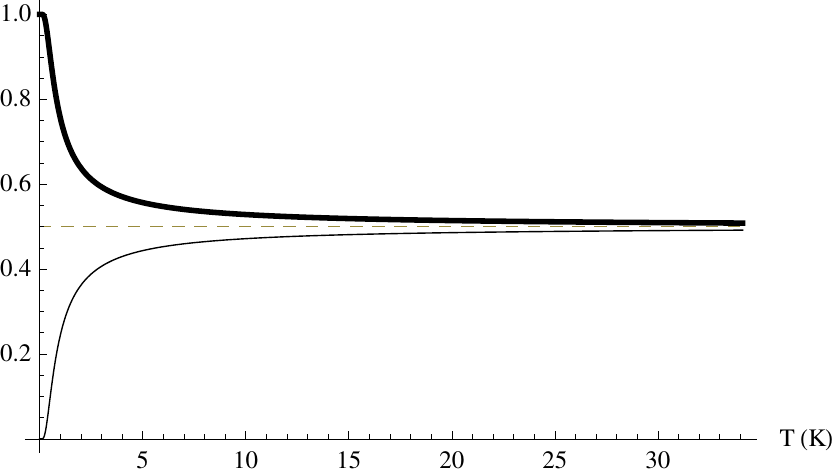}}
\caption{
The fraction of molecules in the first (thick line) and second (solid line) eigen-energies of Ammonia (as an example of a molecule with double-well potential) as a function of temperature (K). The partition function is obtained from the spectroscopic data of Ammonia in [12].
}
\end{center}
\end{figure}

\section{The effect of intermolecular interactions}. Several mechanisms have been proposed to describe how intermolecular interactions affect the dynamics of chiral molecules \cite{HS,SH,Sim,Pf,JZ,Wig,TH,AK,Ber}. Here we focus on interactions due to collisions (also known as the collisional decoherence), which also is very common in nature. In the two-state approximation, the collisional decoherence master equation is given by a Lindblad-type master equation \cite{HS,SH,Sim,Pf,JZ,TH}:
\begin{equation}
\label{eq:eq7}
\frac{\partial \hat{\rho}}{\partial t} =
-i[\hat{H_0},\hat{\rho}]
-\frac{\lambda}{2}\left(\hat{\rho}-\hat{\sigma}_{z}\hat{\rho}\hat{\sigma}_{z}\right),
\end{equation}
where $\lambda$ is the decoherence rate.
An important property of Eq.~\eqref{eq:eq7} is the dynamical transition of an initially delocalized superposition state (e.g., $\psi_{1,2}$) into an incoherent mixture of $\psi_{L}$ and $\psi_{R}$ when $\lambda\gg\omega_x,\omega_z$~\cite{HS,SH,Sim,Pf,JZ,TH}, thus results to racemization, sometimes called as racemization by dephasing~\cite{SH}. However, the dephasing racemization should be distinguished from the typical thermal racemization. In thermal racemization the molecules are energetic enough to surmount the barrier $V_0$, while the incoherent tunneling through the barrier induced by intermolecular interactions results to the dephasing racemization.
We should remind that the decoherence process does not induce the chiral selectivity, i.e., it cannot be used to separate a racemic mixture into its chiral components or it does not result the enantiomeric excess if we start from a racemic mixture.

The great challenge is determination of $\lambda$ which depends on the details of the intermolecular interactions. For example, in \cite{TH} the collisional decoherence effect on $\text{D}_2\text{S}_2$ molecules by Helium atoms as the bath particles has been numerically computed where they considered the chirally sensitive EQ-ED/ED-ED London dispersion interaction between a chiral and an achiral molecule.
Here we use a more general and simple method, first introduced in~\cite{BB}, to compute the collisional decoherence rate for non-planar molecules in gas medium.

The influences of intermolecular interactions on dynamics of chiral molecule can be treated by a kinetic description based on interaction events that can be described as scattering processes. Thus, we may apply the theory of quantum Brownian motion \cite{VH} to describe the effects of collisional interactions. An important observation is that $\psi_{1,2}$ in a chiral molecule are spatial superpositions of an atom or a group of atoms between two wells of double-well potential. We may consider the chiral molecule as a Brownian particle in a superposition of distinct spatial positions separated by the distance $q_0$. The Brownian particle interacts with a bath of scattering particles. We work in dilute gas limit where three-body collisions or correlated scatterings are negligible. Accordingly, when the scattering is recoil-free and also isotropic, the decoherence rate can be obtained as follows \cite{Adl,VH,Dios,BB}:
\begin{equation}
\label{eq:eq8}
\lambda\; = \; \Gamma_t-
n\int_{0}^{\infty} dp \: \nu\left(p\right) \frac{p}{m} \sigma(p,q_0)
\end{equation}
where
$\Gamma_t$ is the total collision rate,
$m$ the mass of bath particles,
$n$ the density of bath particles,
$p$ the bath particle momentum,
$\nu\left(p\right)$ the distribution of such momenta,
$f\left(p,\theta\right)$ the scattering amplitude,
$\theta$ the scattering angle, and
\begin{equation}
\label{eq:eq9}
\sigma(p,q_0) = 2\pi \int^{+1}_{-1} d\left(\cos \theta\right)
\frac{\sin \Theta}{\Theta}
\left|f\left(p,\theta\right)\right|^{2}
\end{equation}
with
$\Theta=(2pq_{0}/\hbar)\sin\left(\theta/2\right)$.
We assume that, for all important bath particles' momenta $p$, the $\sigma\left(p,q_{0}\right)$ does not change appreciably. We can then write:
\begin{equation}
\label{eq:eq81}
\lambda\; \simeq \; \Gamma_t-n \frac{\bar{p}}{m} \sigma(\bar{p},q_0)
\end{equation}
with $\bar{p}=\sqrt{2mk_B T}$ the most probable momentum of an ideal gas.
As custom, we can distinguish limiting cases of Eq.~(\ref{eq:eq81}).
In the high-temperature limit where $\bar{p}q_0 \gg \hbar$, we have $\sin\Theta/\Theta \simeq 0$ \cite{Adl,VH,Dios}. Then, for the decoherence rate one obtains:
\begin{equation}
\lambda \simeq \Gamma_t, \qquad\qquad\qquad T \gg \frac{\hbar^{2}}{2m k_{B}q_{0}^{2}}
\end{equation}

For $q_0$ of the order of few Angstrom (i.e., molecular ranges), and $m\approx4-30 \,\text{gr/mol}$ (from Helium to air), one gets: $\frac{\hbar^{2}}{2m k_{B}q_{0}^{2}}\approx10^{-2}-10\:\text{K}$. Accordingly, in normal conditions, a good estimation can be obtained if we consider the mean rate of collisions as a measure for the order of magnitude of decoherence rate.

The total collision rate is determined by the total scattering cross-section. By exploiting the optical theorem, the total scattering cross-section is connected to the imaginary part of scattering amplitude $f\left(p,\theta\right)$ ~\cite{Mes,Joach}. Using the phase-shift methods, $f\left(p,\theta\right)$ can be determined as a function of interaction potential~\cite{Mes,Joach}. A simple and general expression of the total collision rate can be obtained in the hard-sphere approximation \cite{Mor,Mes,Joach}:
\begin{equation}\label{eq:eq11}
\Gamma_t \simeq \frac{P}{k_B T} (r_1+r_2)^2\left(\frac{8\pi k_B T}{m}\right)^{1/2}
\end{equation}
where $P$ is the pressure of bath particles,
$r_1$ is the effective hard-sphere of bath particle
and $r_2$ is the effective hard-sphere of chiral molecule.

We now investigate the regime in which the molecular parity violation is greater than the dephasing racemization effects of an environment. To give an estimate, let us consider a chiral molecule in the air.
Noting that at best $\omega_z \simeq 3\times10^{-2}$Hz for heavy chiral molecules where recoil-free assumption is valid (i.e., $m_\text{\tiny{chiral molecule}} \gg m_\text{air}$), $r_2$ has the range of a few Angstrom \cite{Mor}, $r_1 \simeq a_{\text{N}_2}$ is the effective hard-sphere of Nitrogen gas, we find from Eq.~\eqref{eq:eq11} that in the limits:
\begin{eqnarray}
\frac{P(\text{Pa})}{\sqrt{T(\text{K})}}&\leq&
3\times10^{-2}\frac{\sqrt{ k_B m_\text{air}}}
{2 (a_{\text{N}_2}+r_2)^2\sqrt{2\pi}}
\\
\frac{P(\text{Pa})}{\sqrt{T(\text{K})}}&\leq&(10^{-8}-10^{-7})
\end{eqnarray}
the parity-violating interactions are physically important (i.e., $\omega_z\geq\lambda$) and they control the dynamics of chiral molecules. It is clear that at the normal conditions ($P=1$atm and $T=300$K), the decoherence rate is much greater than the strongest molecular parity-violation ($\lambda_{P=1\text{atm},\,T=300\text{K}}\approx 10^9-10^{10}$Hz) and so the environmental effects smear out the effects of parity-violating interactions. Introducing the temperature of few Kelvins, we obtain the pressure should be less than $\sim10^{-7}$Pa.
These conditions are in a very good agreement with the experimental conditions in supersonic beam molecular spectroscopy which is the main technique that is being used, aiming at measuring the molecular parity violation~\cite{Dar}.

\section{Concluding remarks}.
We show that how parity-violating interaction can stabilize the optical activity of an isolated molecule. Accordingly, for some molecules the parity violating interactions can partially resolved the problem of optical activity. However, the question remains open why the chiral molecules have not been observed in a superposition of their chiral states. To find a proper explanation, we should also include the environmental interactions (in particular collisions with a background gas) into the dynamics. The environment may induce the dynamical destruction of the quantum coherence, thus results to the racemization by dephasing. The important issue is to find the experimental conditions in which the parity-violating interactions are physically important, e.g., they are greater than racemization effects of intermolecular interactions. Considering the quantum coherence in chiral molecules as a spatial superposition, we used the quantum Brownian motion to compute the conditions in which the parity violating effects may have physically observable effects. However, here we used the simplest representation of the interaction potential (i.e., hard sphere). More sophisticated model potential can be used to calculate the scattering amplitude, and so to arrive at more accurate quantitative results.

The above calculations may also shed light on the conundrum of the origin of life and particularly origin of homo-chirality.  There are some suggestions that the molecular parity-violating energies (although not yet observed experimentally) may have possibly important consequences for the question of bio-homochirality, and thus the origin of life \cite{chi1,chi2,chi3,biochi1,biochi3,B,Q1,Q2}. However, much more theoretical and experimental work is needed to find out whether or not there is any connection between parity violation and bio-homochirality because in these studies the chiral molecules should be considered as an open quantum system interacting with its environment.

\textbf{Acknowledgment}: M.B. acknowledges the partial financial support from Iran National Elite Foundation, and the hospitality of Dr. A. Bassi and the Condense Matter and Statistical Physics section of International Center of Theoretical Physics, Trieste, Italy, where the original idea of this work was developed.

\end{document}